\documentclass[doublecol]{epl2}

\usepackage{ifthen}
\usepackage{ifpdf}

\usepackage{latexsym}
\usepackage{amsmath}
\usepackage{amssymb}
\usepackage{bm}
\usepackage{wasysym}

\ifpdf
\usepackage{graphicx}
\usepackage{epstopdf}
\else
\usepackage{graphicx}
\usepackage{epsfig}
\fi

\newcommand{\trc}{\mbox{trace}}

\newcommand{\im}{\mbox{Im}}

\newcommand{\eexp}{\mbox{e}^}

\newcommand{\tbox}[1]{\mbox{\tiny #1}}

\newcommand{\amatrix}[1]{\begin{matrix} #1 \end{matrix}}

\newcommand{\beq}[1]{\begin{eqnarray}\ifthenelse{#1=-1}{\nonumber}{\ifthenelse{#1=0}{}{\label{e#1}}}}
\newcommand{\eeq}{\end{eqnarray}}

\newcommand{\hide}[1]{}

\newcommand{\rr}{{\bf r}}
\newcommand{\RR}{{\bf R}}
\newcommand{\qq}{{\bf q}}

\title{Decoherence of a particle in a ring}
\shorttitle{Decoherence}

\author{Doron Cohen and Baruch Horovitz}

\institute{Department of Physics, Ben Gurion university, Beer Sheva 84105 Israel}


\abstract{
We consider a particle coupled to a dissipative environment and
derive a perturbative formula for the dephasing rate based on the purity
of the reduced probability matrix. We apply this formula
to the problem of a particle on a ring,
that interacts with a dirty metal environment.
At low but finite temperatures we find a dephasing rate $\propto T^{3/2}$,
and identify dephasing lengths for large and for small
rings. These findings shed light on recent Monte Carlo data
regarding the effective mass of the particle.
At zero temperature we find that spatial fluctuations
suppress the possibility of having a power law decay of coherence.
}

\begin{document}

\maketitle


\section{Introduction}

The problem of dephasing of a particle coupled to a dissipative
environment at temperature $T$, and in particular in the limit
$T\rightarrow0$ has fascinated the mesoscopic community during the
last two decades \cite{AAK,imry,webb,zaikin,alt,golub,delft}. It
has been shown \cite{dld,qbm} that the Caldeira-Leggett (CL)
framework \cite{FV,CL} can be generalized and that the proper way
to characterize the environment is by its form factor
$\tilde{S}(q,\omega)$. Application of the Feynman-Vernon formalism
\cite{dld,qbm} and a semiclassical analysis have shown that an
interference amplitude $P_{\varphi}$ decays with time
as $P_{\varphi}=\exp(-p_{\varphi}(t))$ with
\beq{1}
p_{\varphi}(t) \ \ = \ t \int_q \int_{\omega}
\, \tilde{S}(\bm{q},\omega) \, \tilde{P}(-\bm{q},-\omega)
\eeq
where the integration measures over the wavevector and the
frequency are $d^3q/(2\pi)^3$ and $d\omega/2\pi$ respectively.
The interference suppression factor $P_{\varphi}$ is known 
in the literature as the {\em dephasing} \cite{imry} 
or as the {\em decoherence} factor \cite{weiss}, 
and in the present work we show that it reflects loss of {\em purity}.
In the semiclassical treatment $\tilde{S}(q,\omega)$ is the {\em
symmetrized} form factor of the environment and
$\tilde{P}(q,\omega)$ is the classical {\em symmetric} power
spectrum of the motion.
It has been conjectured using ad-hoc argumentation \cite{dph} (see
also \cite{imry,florian}) that ``the correct" procedure is to use
non-symmetrized spectra. One of our aims is to provide a proper
derivation for a corrected Eq.(\ref{e1}).

During the last decade the study of a particle in a ring coupled
to a variety of environments, has become a paradigm for the study
of ground state anomalies
\cite{Buttiker1,florian2,guinea,golubev,horovitz}. Besides being a
prototype model problem it may be realized as a mesoscopic
electronic device, and it is also of relevance to experiments with
cold atoms or ions that are trapped above an ``atom chip" device
\cite{harber,jones,lin}, where noise is induced by nearby metal
surfaces.
A significant progress has been achieved in analyzing the
equilibrium properties of this prototype system, in particular the
dependence of the ground state energy on the Aharonov Bohm flux
through the ring.

In the present work we define ``dephasing" as the {\em progressive
loss of purity} and find a consistent revised form of
Eq.(\ref{e1}) that is valid beyond the semiclassical context. We
apply this result to the model of a particle on a clean ring that
interacts with a dirty metal environment. At finite temperature we
identify the dephasing rate  ${\Gamma_{\varphi} =
p_{\varphi}(t)/t}$, that vanishes at zero temperature. At $T{=}0$
we find that {\em only} in the CL-like limit of our model there is
still slow progressive spreading (${p_{\varphi}(t)\sim \ln t}$)
which suggests a power law decay of coherence.
Our results shed new light on recent
Monte Carlo data for the temperature dependence of
mass-renormalization \cite{horovitz2}.


\section{Purity}

Our starting point is the most natural definition for the
dephasing factor as related to the purity $\trc(\rho^2)$ of the
reduced probability matrix. The notion of purity is very old, but
in recent years it has become very popular due to the interest in
quantum computation \cite{purity}. Assume that the state of the
system including the environment is $\Psi_{pn}$, where $p$ and $n$
label the basis states of the particle and the bath respectively.
Tracing the environment  states $n$ defines a reduced probability
matrix $[\rho_{sys}]_{p,p'}=\sum_n \Psi_{pn}\Psi_{p'n}^*$ and the
purity is then measured by the dephasing factor
$P_{\varphi}=\sqrt{\trc(\rho_{sys}^2)}$.
Assuming a factorized initial preparation as in the conventional
Feynman-Vernon formalism, we propose the loss of purity
(${P_{\varphi}<1}$) as a measure for decoherence. A standard
reservation applies: initial transients during which the system
gets ``dressed" by the environment should be ignored as these
reflect renormalizations due to the interactions with the high
frequency modes. Other choices of initial state might involve
different transients, while the later slow approach to equilibrium
should be independent of these transients. In any case the
reasoning here is not much different from the usual ideology of
the Fermi golden rule, which is used with similar restrictions to
calculate transition rates between levels.

Consider then a factorized initial preparation
${\Psi^{(0)}_{pn}=\delta_{p,p_0}\delta_{n,n_0}}$, so that within
perturbation theory all $\Psi_{pn}$ are small except for
$\Psi_{p_0n_0}$. We can relate $P_{\varphi}$ to the probabilities
${P_t(p,n|p_0,n_0)=|\Psi_{pn}|^2}$ to have a transition from the
state $|p_0,n_0\rangle$ to the state $|p,n\rangle$ after time~$t$.
To leading order we find
\beq{2}
P_{\varphi} &=& P_t(p_0,n_0|p_0,n_0) \\
&+& \sum_{p\neq p_0}P_t(p,n_0|p_0,n_0)
+ \sum_{n\neq n_0}P_t(p_0,n|p_0,n_0) \nonumber
\eeq
The first term in Eq.(\ref{e2}) is just the survival probability
$P{\tbox{survival}}$ of the preparation.  The importance of the
two other terms can be demonstrated using simple examples: For an
environment that consists of static scatterers we have
$P_{\tbox{survival}}<1$ but $P_{\varphi} = 1$ thanks to the second
term. For a particle in a ring that interacts with a $q=0$
environmental mode $P_{\tbox{survival}}<1$ but $P_{\varphi} = 1$
thanks to the third term.
Using $\sum_{p,n}P_t(p,n|p_0,n_0)=1$ we finally obtain
\beq{3}
p_{\varphi} \ \ = \ \ 1-P_{\varphi} \ \ = \ \ \sum_{p\neq p_0}\sum_{n\neq n_0}P_t(p,n|p_0,n_0)
\eeq
This result has the form of a Fermi-golden-rule (FGR), i.e. it is
the probability that both the system and the bath make a
transition. This differs from the usual FGR treatment \cite{imry}
in which terms like $P_t(p_0,n\neq n_0|p_0,n_0)$ are included.
In the problem that we consider in this paper we can calculate
$P_{\varphi}$ using a $dq d\omega$ integral as in Eq.(\ref{e1}).
In many examples the $\omega=0$ transitions have zero measure and
therefore $P_{\varphi}$ is practically the same as
$P_t(p_0,n_0|p_0,n_0)$. Otherwise one has to be careful in
eliminating those transitions that do not contribute to the
dephasing process. Anticipating the application of Eq.(\ref{e1})
for the calculation of the dephasing for a particle in a ring, the
integration over~$q$ becomes a discrete summation where the
$q{=}0$~related component should be excluded.
It is implicit in the derivation of Eq.(\ref{e1}) from Eq.(\ref{e3})
that at the last step a thermal average is taken over
both $n_0$ and $p_0$, though in general one may consider non-equilibrium
preparations as well.


\section{Dephasing formula}

We would like to apply our revised FGR Eq.(\ref{e3}) to the
general problem of a particle at position $\RR$ coupled to an
environment with electronic density $\mathsf{n}(\rr,t)$. It is
implicit that the particle also experiences an external potential
that defines the confining geometry. A Hamiltonian
${\cal H}_0$ of the particle in the confined geometry defines the
states and eigenstates via ${\cal H}_0|p\rangle=E_p|p\rangle$.
For definiteness we use the Coulomb interaction, though any other
interaction may be used, hence the particle -
environment interaction is
\beq{4} V_{\tbox{int}} = \int d^3r \rho(\rr) \int d^3r' \frac{e^2
\mathsf{n}(\rr',t) }{|\rr-\rr'|} \equiv \int d^3r \rho(\rr) \,
\mathcal{U}(\rr) \eeq
where $\rho(\rr) =  \delta(\rr-\RR(t))$, with $\RR(t)$ the
position operator of the particle in the Heisenberg (interaction)
picture. Our FGR with $P_{t}(p,n|p_0,n_0)=|\langle
p,n|\int_0^{t}V_{\tbox{int}}dt'|p_0,n_0\rangle |^2$
(using $\hbar=1$ units) yields
\beq{30} p_{\varphi} &=& e^2 \sum_{p (\neq p_0)} \sum_{n (\neq
n_0)} \int_0^{t}dt'\int_0^{t}dt''\int_r\int_{r'} \nonumber
\\
&&
\langle p_0 | \rho(\rr'',t'')  | p \rangle
\langle p | \rho(\rr',t')  | p_0 \rangle \,
\nonumber
\\
&&
\langle n_0|\mathcal{U}(\rr'',t'')|n\rangle
\langle n|\mathcal{U}(\rr',t')| n_0\rangle
\label{e5}
\eeq
The double time integral can be written
as a $dq d\omega$ integral over Fourier components.
For this purpose we define the form factor
of the {\em fluctuations} (as seen by the particle):
\beq{8}
\tilde{S}(\bm{q},\omega)
=\int d^3r\int d\tau
\,\langle \mathcal{U}(\rr',t')\mathcal{U}(\rr,t)\rangle
\,\eexp{i\omega\tau-i\qq\cdot\rr}
\eeq
with thermal average replacing the $n_0$ state expectation value.
$\tilde{S}(q,\omega)$ is related to the dielectric function of the
environment $\varepsilon(\bm{q},\omega)$ via the fluctuation dissipation
theorem
\beq{9}
\tilde{S}(\bm{q},\omega) =
\frac{4\pi e^2}{\bm{q}^2} \im\left[ \frac{-1}{\varepsilon(\bm{q},\omega)} \right]
\frac{2}{1-\eexp{-\omega/T}}
\eeq
In the semiclassical formulation one replaces the operator
$\bm{R}(t)$ by the classical trajectory $\bm{R}_{cl}(t)$, and
consequently in Eq.(\ref{e5}) the particle-related part of the
integrand is replaced by a classical two-point correlation
function of the type $\langle f(\bm{R}_{cl}(t''))
f(\bm{R}_{cl}(t')) \rangle$. In the quantum context the particle
dependent part of Eq.(\ref{e30}),  after Fourier transform, leads
to the following definition for the power spectrum of the motion:
\beq{100}
\tilde{P}(\bm{q},\omega) =
\int
\Big[
\langle \eexp{-i\bm{q}\cdot \bm{R}(\tau)} \eexp{i\bm{q}\cdot \bm{R}(0)} \rangle
- \langle \eexp{i\bm{q} \cdot \bm{R}}\rangle^2
\Big]
\, \eexp{i\omega \tau} \, d\tau
\eeq
Also here, close to equilibrium condition, a thermal average
should replace the $p_0$ state expectation value. It is important
to realize that this definition, as well as Eq.(\ref{e9}), imply
that non-symmetrized spectral functions should be used.
Our main interest is in very low temperatures,
so we set for presentation purpose $p_0=0$. Then we get
\beq{10} \tilde{P}(\bm{q},\omega)= \sum_{p\neq 0} |\langle
p|\eexp{i\qq\cdot\RR}|0\rangle|^2 \, \delta(\omega-E_p) \eeq
Using the above {\em definitions} for  $\tilde{S}(\bm{q},\omega)$
and $\tilde{P}(\bm{q},\omega)$ we can re-write
\beq{0}
p_{\varphi} = \int_0^{t}dt'\int_0^{t}dt'' \int_q
\int\!\!\!\!\!\int_{\omega',\omega''} \tilde{S}(\bm{q},\omega')
\tilde{P}(-\bm{q},-\omega'') \,
\nonumber \\
\times\eexp{-i(\omega' -\omega'')(t''-t')}\,.
\label{e11}
\eeq
For practical calculations or for aesthetic reasons we prefer to
use soft rather than sharp cutoff for the time integration. Then
we get
\beq{101} p_{\varphi} = t \int_q
\int\!\!\!\!\!\int_{\omega,\omega'} \!\!\!\!\!
\tilde{S}(\bm{q},\omega) \tilde{P}(-\bm{q},-\omega')
\left[\frac{(2/t)}{(1/t)^2+(\omega{-}\omega')^2}\right] \eeq
This result can be cast into the form of Eq.(\ref{e1}) provided
$\tilde{P}(\bm{q},\omega)$ is {\em re-defined} as the convolution
of Eq.(\ref{e100}) with the kernel in the square brackets, which
is like time-uncertainty broadened delta function. Eq.(\ref{e101})
is our revised form of Eq.(\ref{e1}); it provides the dephasing
factor $P_{\varphi}=\exp(-p_{\varphi})$ for a general
particle-environment interaction.

\section{Dephasing rate}

At finite temperatures, if $t$ is larger compared with dynamically
relevant time scales, and in particular ${t \gg 1/T}$,
we can replace the square brackets in Eq.(\ref{e101}) by $2\pi
\delta(\omega{-}\omega')$. Consequently we get linear growth
$p_{\varphi} \approx \Gamma_{\varphi} t$ with the rate
\beq{12}
\Gamma_{\varphi}
=\int_{q} \int_{\omega}
\tilde{S}(\bm{q},\omega) \tilde{P}(-\bm{q},-\omega)
\eeq
Following standard arguments one conjectures that the long time
decay is exponential, i.e. $P(t)=\exp(-\Gamma_{\varphi}t)$, as in
the analysis of Wigner's decay.
In terms of the dielectric function $\varepsilon(q,\omega)$  we
obtain the following general result:
\beq{13}
\Gamma_{\varphi}
=
\sum_{p\neq 0}
\int_q
\frac{4\pi e^2}{q^2}
\im\left[\frac{1}{\varepsilon(q,{-}E_p)}\right]
\frac{2|\langle p|\eexp{i\qq\cdot\RR}|0\rangle|^2}{\eexp{E_p/T}{-}1}
\eeq
Our assumption ${t\gg(1/T)}$ implies that (\ref{e13}) can be
trusted only if ${\Gamma_{\varphi} \ll T}$ which implies
a weak coupling condition (see below).

\section{Dirty metal}

So far we kept the derivation general, without specifying either
the particle states $|p\rangle$ or the dielectric function
$\varepsilon(q,\omega)$. We consider now a particle of
mass $M$ on a ring of radius $R$ 
so that ${\cal H}_0=-(2MR^2)^{-1}\partial^2_{\theta}$ 
where $\theta$ is the angle variable. 
The particle eigenstates are then $|p_m\rangle \propto\eexp{im\theta}$ 
with energy eigenvalues $E_m=m^2/(2MR^2)$. 
We study the effect of low frequency fluctuations
(${|q|\lesssim1/\ell, |\omega|\lesssim\omega_c}$) due to a dirty
metal environment for which
${\varepsilon(q,\omega)=1+4\pi\sigma(-i\omega+Dq^2)^{-1}}$, where
$\sigma$ is the conductivity, $D$ is the diffusion constant, and
$\ell$ is the mean free path. Below we identify the renormalized
value of the high frequency cutoff~$\omega_c$ as the classical
damping rate $\gamma_r = 2\pi\alpha/M\ell^2$, where the
dimensionless interaction strength is
$\alpha=e^2/(8\pi^2\sigma\ell)=3/(8(k_F\ell)^2)$ and $k_F$ is the
Fermi wavevector.
We first consider the case of a large ring with $r=R/\ell \gg 1$.
Using the Fourier expansion \cite{guinea,golubev}
\beq{14}
&&\ell\int_q
\eexp{-i\qq\cdot(\RR(\theta)-\RR(\theta'))}\frac{4\pi}{q^2}=
\frac{1}{\sqrt{4r^2\sin^2(\frac{\theta-\theta'}{2})+1}}\nonumber\\
&& \ \ \ \ \ \ \ \
= 1-\sum_m a_m \sin^2\left(\frac{m(\theta-\theta')}{2}\right)
\eeq
with
\beq{15}
a_m\approx
\left\{\amatrix{
\frac{2}{\pi r}\ln \frac{r}{m}, & 1\leq m \leq r \cr
0, & \mbox{otherwise}
}\right.
\eeq
we have
\beq{16}
\int_q \frac{4\pi}{q^2}|
\langle0|e^{-i\qq\cdot\RR}|p_m\rangle|^2
=\frac{1}{4} a_{|m|}
\eeq
and therefore
\beq{17}
\Gamma_{\varphi}
=
2\pi\alpha\sum_{m\ne0} a_m\frac{E_m\eexp{-|E_m|/\omega_c}}{\eexp{E_m/T}{-}1}
\approx 2\pi\alpha T
\!\!\!\!\! \sum_{0<|m|<r_{\tbox{eff}}} \!\!\!\!\! a_m
\eeq
where $r_{\tbox{eff}} \equiv \min\{r, (2MR^2T)^{1/2},
(2MR^2\omega_c)^{1/2}  \}$ is determined by the conditions ${m<r}$
and $E_{m} < T$ and $E_m<\omega_c$. At high temperatures,
$T>\omega_c$, $r_{\tbox{eff}}$ is temperature independent and
therefore $\Gamma_{\varphi}\propto T$, while at low temperatures
[but still ${T>1/(2MR^2)}$] we get, as shown schematically in
Fig.~1,
\beq{18}
\Gamma_{\varphi}=4\alpha T\sqrt{2M\ell^2T}|
\ln\sqrt{2M\ell^2T}|\sim T^{3/2}|\ln T|
\eeq
From these results it follows that the self consistency
requirement $\Gamma_{\varphi}\ll T$, as discussed after
Eq.(\ref{e13}), is globally satisfied for any temperature if
$\alpha \ll 1$, or equivalently if ${k_F\ell\gg 1}$; in the regime
of Eq.(\ref{e18}) the constraint $\Gamma_{\varphi}\ll T$ is
satisfied also with stronger $\alpha$.
We note that if the ${m{=}0}$
Fourier component were included in the summation, then the low $T$
form would change to $\Gamma_{\varphi} \propto T$, in contrast
with the proper result Eq.(\ref{e18}).

Though the derivation of Eq.(\ref{e101}) refers to a "one
particle", it turns out that the treatment of dephasing in the
many body problem is not much different. Namely, the effect of the
"Pauli principle" in the low temperature Fermi sea occupation can
be incorporated via an appropriate modification of the cutoff
scheme for the spectral functions involved. On the heuristic level
it involves a~$T$ dependent momentum cutoff as in~\cite{dph},
while more recently it has been formulated and established using
more advanced methods \cite{Munich}.

\begin{figure}
\includegraphics[width=\hsize]{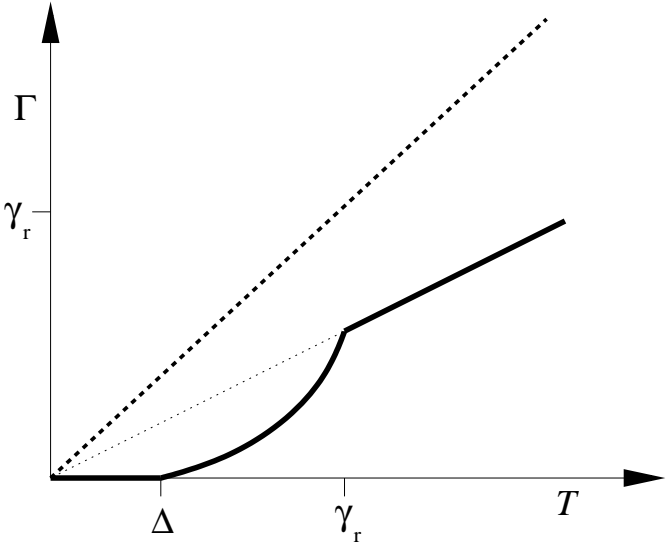}
\caption{Illustration of the dependence of the dephasing
rate~$\Gamma$ on the temperature~$T$.  The dephasing rate is well
defined for ${t>(1/T)}$, and hence the self consistency
requirement is $\Gamma\ll T$. This condition is demonstrated by a
comparison with the dashed line. The illustration assumes weak
coupling $\alpha\ll 1$ and large rings $\alpha r^2\gg 1$ so that
the energy cutoff is $\gamma_r=2\pi\alpha/M\ell^2 \gg \Delta=1/MR^2$.
 For extremely low
temperatures, such that $T$ is smaller compared with the spacing
$\Delta$, the probability to excite the system is exponentially
small and the familiar two-level modeling becomes applicable.}
\label{fig2}
\end{figure}


\section{Zero temperature}

We would like to discuss the ``zero temperature" regime.
If the temperature were extremely low, such that $T \ll \Delta$
where $\Delta \sim 1/(MR^2)$ is the ground state level spacing,
we could treat the problem using a ``two level approximation",
which is a very well studied model \cite{weiss}.
We do not further discuss this regime. From here on
we assume $T \gg \Delta$. Thus we can treat the $d\omega$
integration as if the levels of the ring form a continuum.
But we still can define ``zero temperature" as such for which the
practical interest is in the time interval $t \ll 1/T$, which can
be extremely long.
Then one realizes that Eq.(\ref{e101}) gives a
non-zero result even at ``zero temperature":
\beq{21}
p_{\varphi} \approx
\alpha \sum_m a_m \ln\frac{\omega_c}{E_m+(1/t)}
\eeq
where $\omega_c$ is the high frequency cutoff of the environmental
modes. Assuming ${\ell \ll R}$ we get after a transient:
\beq{36}
p_{\varphi} \ \ \approx \ \ \alpha\ell \int_0^{1/\ell} dq
\,
 \, \ln\left[\frac{1}{q\ell}\right] \,
\ln\left[\frac{M\omega_c}{q^2}\right] \ \ \approx \ \  \alpha \eeq
where for clarity of presentation we converted the $m$~summation
into a ${dq \equiv (1/R) dm}$ integral.
Accordingly we deduce that for ${r \gg 1}$
coherence is maintained if $k_F\ell \gg 1$.
This should be contrasted with the CL limit
(${\ell\rightarrow\infty}$) where the integral has a singular
${q\sim0}$ contribution from the lowest fluctuating mode
(${m=1}$), and consequently ${p_{\varphi} \approx 2\alpha
r^2\log(\omega_ct)}$, which is a well known expression
\cite{golub,florian2}. 
But once $r<1$ the condition for using 
our perturbation result is replaced by ${\alpha r^2 \ll 1}$, 
which is an $\ell$ independent condition.
In this CL limit the quantization of the energy spectrum 
is important and the renormalized cutoff frequency 
becomes ${\omega_c\sim \Delta}$ instead of ${\omega_c\sim \gamma_r}$. 
Accordingly, in the latter circumstances, 
the time during which the log spreading prevails diminishes.


\section{Effective mass}

It can be shown \cite{dld} that the mass renormalization in the
inertial (polaronic) sense is $\Delta M = \eta/\omega_c$. However
in recent works \cite{guinea,golubev,horovitz} the  mass
renormalization concept appears in a new context. The free energy
$\mathsf{F}(T,\Phi)$ of a particle in a ring is calculated, where
$\Phi$ is the Aharonov Bohm flux through the ring. Then the
coherence is characterized by the ``curvature", which is a measure
for the sensitivity to $\Phi$. The curvature can be parameterized
as
\beq{37}
\left.\frac{\partial^2 \mathsf{F}}{\partial\Phi^2} \right|_{\Phi{=}0}
\ \ = \ \
\frac{e^2}{M^*R^2}f(M^{*}R^2T)
\eeq
where in the absence of environment $M^*=M$ is the bare mass of
the particle, and the $T$ dependence simply reflects the Boltzmann
distribution of the energy. In the presence of coupling to the
environment $M^*>M$ and $M^*$ depends on both $\alpha$ and $T$.
At $T=0$, for fixed $\alpha \ll 1$, Monte Carlo data show
\cite{horovitz2} that the ratio $M^*/M$ is independent of the
radius provided $r>r_c$, where $r_c$ is a critical radius. As the
radius becomes smaller compared with $r_c$, the ratio $M^*/M$
rapidly approaches unity.
In the regime of ``large $R$" the mass renormalization effect
diminishes with the temperature and depends on the scaled variable
$RT$, while for ``small $R$" the ratio $M^*/M$ depends on the
scaled variable $R^4T$.
The natural question is whether we can shed some light
on the physics behind this observed temperature dependence.
Making the conjecture that the temperature dependence
of $M^*/M$ is determined by dephasing it is natural
to suggest the following measure of coherence:
\beq{38}
x(T,R)  =
p_{\varphi}\left(t{=}\frac{1}{\Delta_{\tbox{eff}}}\right)
\approx  \frac{\Gamma_{\varphi}}{\Delta_{\tbox{eff}}}
\approx 2\pi\alpha \bar{a} M R^2 T
\eeq
where $\bar{a}$ is an average value of $a_n$. Eq.(\ref{e38})
describes dephasing at the time $t=1/\Delta_{\tbox{eff}}$, where
${\Delta_{\tbox{eff}} \sim r_{\tbox{eff}} \times (MR^2)^{-1} }$ is
the energy scale that characterizes the``effective" transitions;
hence the variable $x$ measures the level sharpness. For a dirty
metal with $\ell \ll R$ the typical value of the Fourier
components is ${\bar{a} \sim 1/r }$ as implied by Eq.(\ref{e15}).
On the other hand for  a dirty metal with $\ell \gg R$ there is
only one effective mode with ${\bar{a} \sim r^2 }$. Accordingly we
get the $RT$ and the $R^4T$ dependence respectively, in agreement
with the Monte Carlo simulations.
The condition $x < 1/2$ can serve as a practical definition for
having coherence. It can be translated either as a condition on
the temperature, or optionally it can be used in order to define a
coherence length that depends on the temperature. The conjecture
is that $M^*/M$ is a function of $x$.


\section{Summary}

In this paper we derive a new perturbative expression for the
dephasing factor $P_{\varphi}(t)$ and apply it to a particle in a
ring coupled to fluctuations of a dirty metal environment. We find
that the dephasing rate vanishes at ${T=0}$. We also define a
coherence criterion that identifies a dephasing length. The latter
diverges as $T^{-1}$ for large radius and as $T^{-1/4}$ for small radius,
in consistency with Monte Carlo data on mass renormalization.
The renormalized mass is an equilibrium property which affects
temporal correlation functions. But we see that it reflects
nonequilibrium features of the dynamics which are expressed
in the dephasing factor calculation. We find this relation between
equilibrium and nonequilibrium scales an intriguing phenomena.


\acknowledgments

We thank Florian Marquardt and Joe Imry for helpful communications.
This research was supported by a grant from the DIP,
the Deutsch-Israelische Projektkooperation.


\end{document}